\documentclass[sn-nature]{sn-jnl}

\usepackage{graphicx}%
\usepackage{multirow}%
\usepackage{amsmath,amssymb,amsfonts}%
\usepackage{amsthm}%
\usepackage{mathrsfs}%
\usepackage[title]{appendix}%
\usepackage{xcolor}%
\usepackage{textcomp}%
\usepackage{manyfoot}%
\usepackage{booktabs}%
\usepackage{dcolumn}
\usepackage{algorithm}%
\usepackage{algorithmic}%

\begin{document}
\title[Electrostatic dipole polarizability and plasmon resonances of multilayer nanoshells]{Electrostatic dipole polarizability and plasmon resonances of multilayer nanoshells}

\author*[1]{\fnm{Luke. C} \sur{Ugwuoke}}\email{lcugwuoke@gmail.com}
\author[1]{\fnm{Mark. S} \sur{Tame}}\email{mstame@sun.ac.za}

\affil*[1]{ \orgdiv{Department of Physics}, \orgname{Stellenbosch University}, \postcode{Private Bag X1, Matieland 7602}, \country{South Africa}} 

\abstract{
We propose a generalized formula for calculating the dipole polarizability of spherical multilayer nanoshells (MNSs) within the long-wavelength approximation (LWA). 
Given a MNS with a finite number of concentric layers, radii, and dielectric properties, embedded in a dielectric medium, in the presence of a uniform electric field, we show that its frequency-dependent and complex dipole polarizability can be expressed in terms of the dipole polarizability of the preceding MNS. 
This approach is different from previous more involved methods where the LWA polarizability of a MNS is usually derived from scattering coefficients. 
Using both finite-element method- and Mie theory-based simulations, we show that our proposed formula reproduces the usual LWA results,
when it is used to predict absorption spectra, by comparing the results to
simulated spectra obtained from MNSs with \emph{n} number of layers up to \emph{n} = 6 layers. 
An iterative algorithm for calculating the dipole polarizability of a MNS based on the generalized formula is presented. 
A Fr\"{o}hlich function whose zeroes correspond to the dipolar localized surface plasmon resonances (LSPRs) supported by the MNS is proposed. 
We identify a pairing behaviour by some LSPRs in the 
Fr\"{o}hlich function that might also be useful for mode characterization.
}


\maketitle 
\section{Introduction}
The optical properties of nanoshells have been studied 
since the works of Kerker \cite{Kerk69}, Wu and Yang \cite{Wu91}, and others \cite{Pena09,Pena13,Pena17}, using Mie theory \cite{BoHu08,Sam07,Mie08}.
The approach presented in this work is a surprisingly simple alternative for obtaining the dipole polarizability of nanoshells using the long-wavelength approximation (LWA) of Maxwell's equations, also known as the quasi-static limit \cite{BoHu08}. Hence, the formulae proposed herein have the usual assumptions associated with the LWA. 
In the LWA, we will show that if the electrostatic polarizability is considered as the physical quantity of interest instead of the ubiquitous approach that is based on scattering coefficients \cite{Far14,Luke22,Soni14,Herr16,Jun14,Ma23,Ma21}, a generalized analytical form of the dipole polarizability can be derived for the $n$th nanoshell given a multilayer nanoshell (MNS) with $n$ number of concentric layers. In addition, a function that predicts the localized surface plasmon resonances (LSPRs) supported by the $n$th nanoshell, which we will refer to as the \emph{Fr\"{o}hlich function}, can be derived using this general form of the dipole polarizability of the $n$th nanoshell when the Fr\"{o}hlich condition \cite{Raza14,Wub14,Sam07} is applied to the polarizability. 

Several authors have shown that the dipole polarizability of a MNS can be calculated using the LWA --- an approximation that leads to the de-coupling of the electric and magnetic field intensities and allows retardation effects such as radiation damping and dynamic depolarization to be ignored, as long as the sizes of the nanoparticles being considered are within the Rayleigh regime \cite{Ma21,Wu06,Herr16,Luke22,Ma23,Radi20,Neev89,BoHu08}. This regime considers particle sizes that are small compared to the excitation wavelength, usually within one-tenth of the excitation wavelength or less \cite{Man06,Raza14, BoHu08,Sam07,Wub14}. 
However, at such sub-100 nm particle sizes, usually around 10 nm or less, metal nanoparticles (MNPs) have been shown to display non-local effects \cite{Kreig85,Raza14,Wub14}. These effects, which are mostly due to increased scattering of the quasi-free electrons near the metal surface, lead to a dependence of the dielectric function of the metal on the longitudinal wavevector of the incident field \cite{Raza14,Wub14}.  In most studies, this dependence is ignored via the introduction of the local response approximation (LRA) \cite{Zhao12,Man06,Soni14}. While the non-local response leads to spectral broadening and suppression of the scattered intensity due to the increased plasmon damping as well as size-induced LSPR shift, it predicts the same number of LSPRs supported by a given MNP when compared to the LRA \cite{Kreig85,Raza14,Wub14}. 

Dielectric-metal/semiconductor or metal/semiconductor-dielectric core-shell nanostructures that are cylindrically- or spherically-symmetric \cite{BoHu08,Pal11,Prodan04,Far14, Radi20,ZaZa13,Ma23} are the building blocks of MNSs --- nanoshells with two or more layers \cite{Ma21,Bard10,Soni14,Zhao12,Herr16,Jun14,Moud14,Nia14,Zhu11}. These nanoshells can be concentric or non-concentric depending on whether the core and shell(s) share a common centre \cite{Bard10,Soni14} or not \cite{Wu06,Luke22,Nort16}. Nanoshells can support one or more LSPRs in their scattering or absorption spectra depending on the material composition \cite{Zhao12,Herr16,Nia14}, sizes of the core and shells \cite{Hu08,Jun14,Zhu11}, or via geometrical symmetry-breaking \cite{Wu06,Luke22,Nort16}. In a MNS, each of the LSPR is either due to a bonding, an anti-bonding, or a non-bonding mode, formed due to the plasmon hybridization of the solid or cavity plasmons of the core with the nanoshell plasmons \cite{Jun14,Hu08,Luke22,Bard10}. 

Previously, Daneshfar and Bazyari \cite{Far14} had proposed a generalized electrostatic dipole polarizability for an $n-$layer nanoshell that depends on the scattering coefficients of the electric fields in the core, shell, and medium regions of the nanoshell, given the material properties, and sizes of the core and shells, which they used to calculate the spectral properties of a MNS with $n=3$ shells. However, Wu and Yang \cite{Wu91} were the first to propose and implement a recursive algorithm for calculating the 
multipole optical response of an $n$-layered sphere in terms of the scattering coefficients of the electric and magnetic field intensities in the core, shell, and host regions of the nanoshell, based on Mie theory \cite{Mie08}, following the works of Kerker \cite{Kerk69}. Pal et al. \cite{Pena09,Pena17,Pena13} improved Yang's work and have used it to predict the dipolar extinction efficiencies of a MNS with $n=3$ layers \cite{Pena17}.
The approach by Wu and Yang \cite{Wu91}, Pal et al. \cite{Pena09,Pena17,Pena13}, and more recent work \cite{Rock18,Pave20,Ivan23,Lee18,Terek17} should 
always be the standard, especially because they account for retardation effects and multipolar response. However, in the LWA, these approaches \cite{Far14,Herr16,Luke22,Ma23,Radi20}---where the dipole polarizability of the $n$th nanoshell is obtained from scattering coefficients---
are not necessary if the polarizability of the preceding $(n-1)$th nanoshell is known, as we will show by reproducing the usual LWA results using the proposed formula.
Our model is less complicated to work with, and
also very insightful, especially when it is used to predict LSPRs via the proposed Fr\"{o}hlich function.

Given the electrostatic dipole polarizability of a MNP, its imaginary part is proportional to the absorption efficiency of the MNP \cite{Wu06,Prodan04}. We will show that the dipole polarizability of a MNS with $n$ shells can be calculated from the dipole polarizability of the preceding $(n-1)$th shell and the material properties and size of the $n$th shell. 
Here, we will employ both the LWA and the LRA to predict the dipolar LSPRs and the absorption spectra of concentric MNS with $n$ layers or shells up to $n = 6$ shells. 
We will show that the formula is more straightforward to work it, and that it reduces to the dipole polarizability of a spherical MNP when there are no shell(s), i.e., when $n = 0$. 
We will also show that the implemented iterative algorithm for predicting the dipole polarizability of the MNS reproduces the usual LWA results by comparing our model to Mie theory-based simulations using the open source python software, scattnlay \cite{Pena09,Pena17,Pena13}, 
as well as finite-element method-based, classical electrodynamics simulations in 3D using Comsol Multiphysics$^{\circledR}$ \cite{comsol,Grand19}. 

\section{Theory and Simulations}\label{sec2}
\begin{figure}[t]
	\includegraphics[width = .10\textwidth]{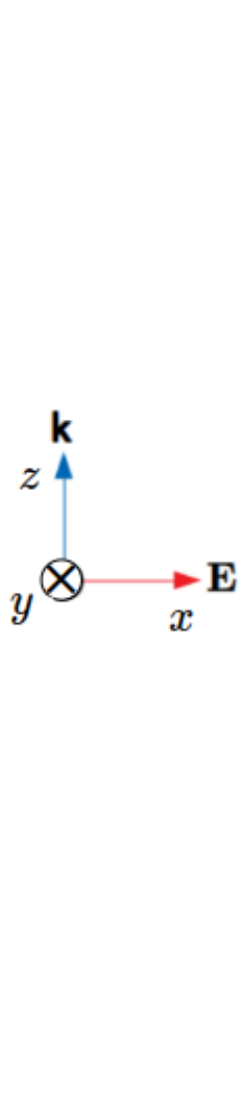}~~~~~~~
	\includegraphics[width = .40\textwidth]{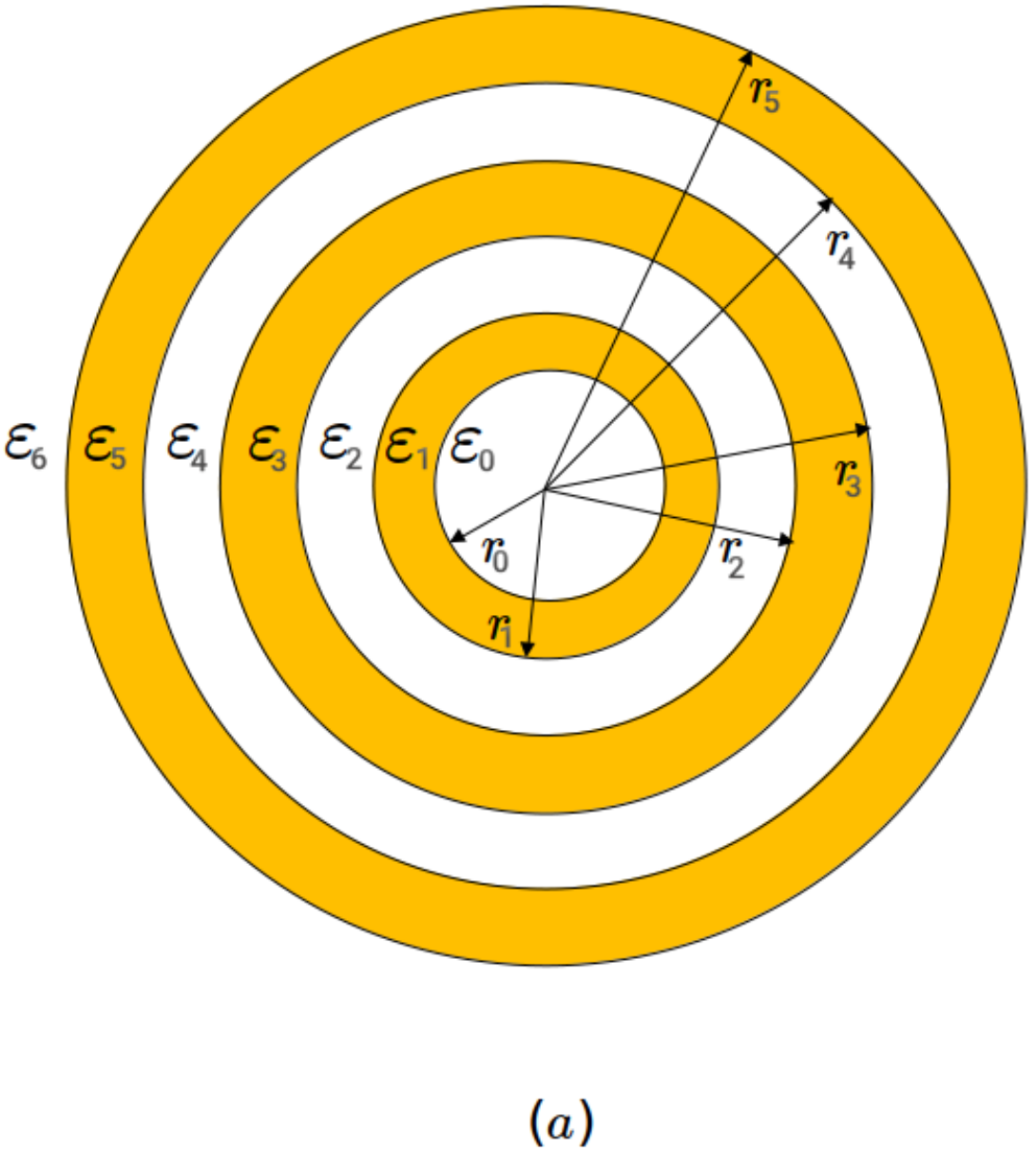}~~~
	\includegraphics[width = .44\textwidth]{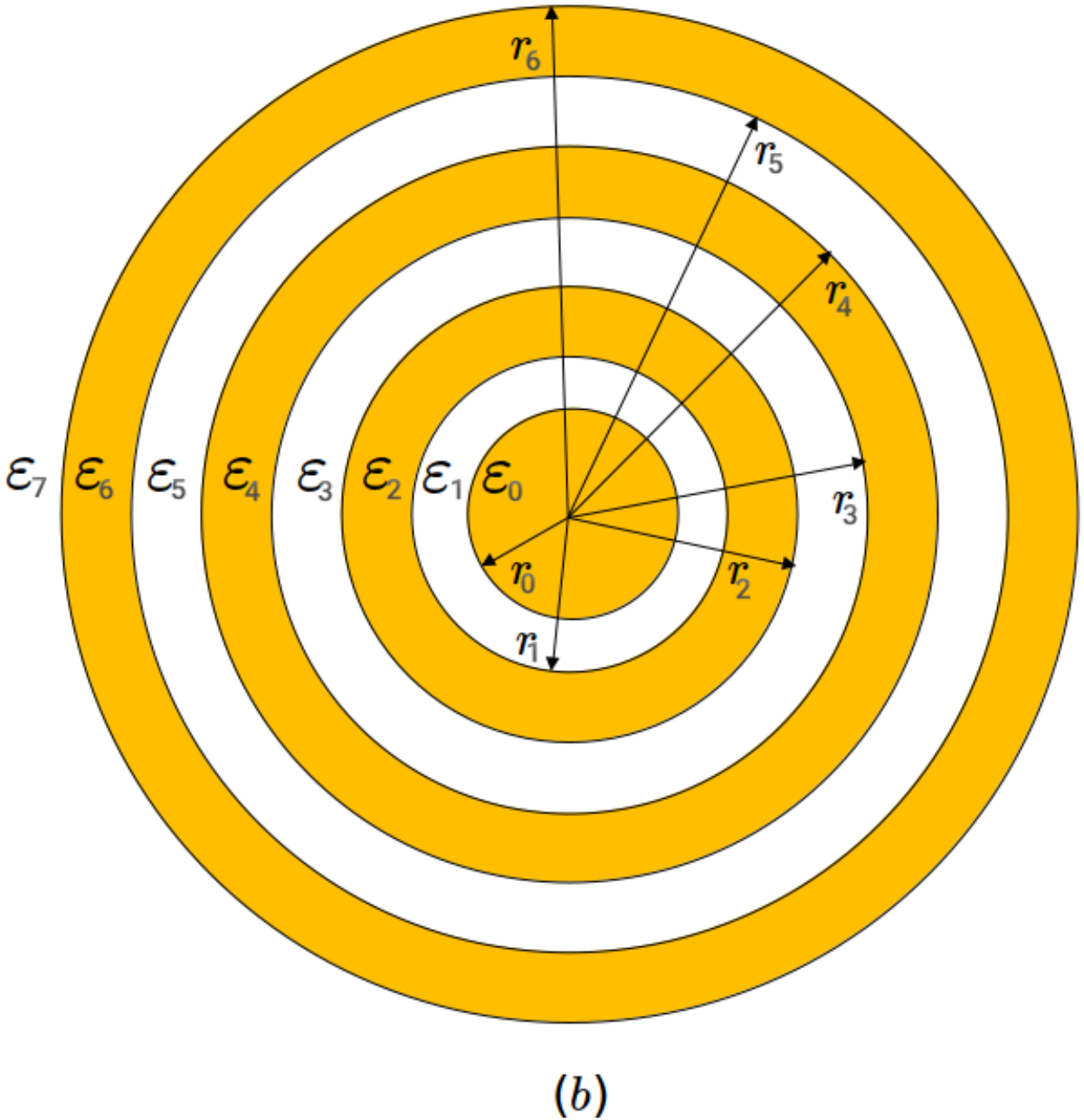}
	\caption{2D schematic of the model geometries of a MNS with: ($a$) 5 layers and ($b$) 6 layers, in the presence of an $x-$polarized incident electric field, $\mathbf{E}$, propagating along the $z-$direction with a wavevector $\mathbf{k}$. The permittivity of the core is $\epsilon_{0}$, that of the medium surrounding the: ($a$) $5$th shell is $\epsilon_{6}$, and ($b$) $6$th shell is $\epsilon_{7}$, and $\epsilon_{n}, n = 1, 2, 3, ..., 6$, are the shell permittivities. The radius of the core is $r_{0}$, and $r_{n}, n = 1, 2, 3, ..., 6$, are the nanoshell radii.
		As we will explain in detail in Section \ref{sec3}, we only consider two kinds of MNSs---in ($a$), the core is a dielectric and in ($b$), the core is a metal. These are the two most common MNS configurations based on material composition but the model is also valid when semiconductors are used.
	}\label{f1}
\end{figure}

\subsection{Electrostatic dipole polarizability}\label{ss1}
An $x-$polarized electric field, $\mathbf{E} = E_{0}e^{-ikz}\hat{e}_{x}$, with amplitude, $E_{0}$,
is incident on the MNS as shown in Fig. \ref{f1}. The magnitude of the wavevector is $k = 2\pi\sqrt{\epsilon_{n+1}}/\lambda$, where $\varepsilon_{n+1}$ is the permittivity of the medium surrounding the $n$th shell, and $\lambda$ is the excitation wavelength. Since $z = r\cos\theta$, given that $r_{n}$ is the radius of the $n$th shell, and $\theta$ is the polar angle, the electric field in each region of a MNS depends on the size parameter, $kr_{n}$, as shown in Refs. \cite{Pena09,Pena17}. However, in the LWA, this is not the case, since $r_{n} << \lambda$, and $kr_{n} \longrightarrow 0$, leading to the following electrostatic dipole polarizability of a sphere (i.e., no shells, $n = 0$) \cite{BoHu08,Sam07}: 
\begin{equation}\label{e1}
	\alpha_{0} = r_{0}^{3}  \left ( \frac{\epsilon_{0} - \epsilon_{1} }{\epsilon_{0} + 2\epsilon_{1} } \right), 
\end{equation}
where Eq. \eqref{e1} is the dipole polarizability normalized by $4\pi\epsilon_{1}$, and $\epsilon_{0}$ and $\epsilon_{1}$ are the permittivities of the sphere and the medium surrounding the sphere, respectively, and $r_{0}$ is the sphere radius. 
When there is only one concentric shell ($n = 1$), the dipole polarizability of the nanoshell in the LWA, derived in Ref. \cite{BoHu08}, and normalized by $4\pi\epsilon_{2}$, is obtained as
\begin{equation}\label{a1}
	\alpha_{1} = r_{1}^{3}
	\frac{\beta_{1} }{\varrho_{1} },
\end{equation}
where
\begin{subequations} 
	\begin{align}
		\beta_{1} & = (\epsilon_{1} - \epsilon_{2})(\epsilon_{0} + 2\epsilon_{1}) + (r_{0}/r_{1})^{3}(\epsilon_{0} - \epsilon_{1})(\epsilon_{2} + 2\epsilon_{1}), \label{a7}\\
		\varrho_{1} & = (\epsilon_{1} + 2\epsilon_{2})(\epsilon_{0} + 2\epsilon_{1}) + 2(r_{0}/r_{1})^{3}(\epsilon_{1} - \epsilon_{2})(\epsilon_{0} - \epsilon_{1}). \label{a8}
	\end{align}
\end{subequations}
Equation \eqref{a1} can be re-written as:
\begin{equation}\label{e2}
	\alpha_{1} = r_{1}^{3}  \left ( \frac{\epsilon_{1} - f_{0}\epsilon_{2} }{\epsilon_{1} + 2f_{0}\epsilon_{2} } \right), 
\end{equation}
where $r_{1}$ is the nanoshell radius, $\epsilon_{1}$ is the permittivity of the shell, $\epsilon_{2}$ is the permittivity of the medium surrounding the shell, and 
\begin{equation}\label{e3}
	f_{0} = \left(1 - \frac{\alpha_{0}}{r_{1}^{3}} \right)\left(1 + 2 \frac{\alpha_{0}}{r_{1}^{3}} \right)^{-1}. 
\end{equation}
When there are two concentric shells ($n = 2$), the dipole polarizability of the nanoshell in the LWA, as derived in Refs. \cite{Luke22,Herr16,Fang08}, and normalized by $4\pi\epsilon_{3}$, is 
\begin{equation}\label{a6}
	\alpha_{2} = r_{2}^{3} \left[
	\frac{\epsilon_{2} \Big(\varrho_{1} + 2(r_{1}/r_{2})^{3}\beta_{1} \Big) - \epsilon_{3}\Big(\varrho_{1} - (r_{1}/r_{2})^{3}\beta_{1} \Big) }
	{\epsilon_{2}\Big(\varrho_{1} + 2(r_{1}/r_{2})^{3}\beta_{1} \Big) + 2\epsilon_{3}\Big(\varrho_{1} - (r_{1}/r_{2})^{3}\beta_{1} \Big) }\right],
\end{equation}
and can be re-written as:
\begin{equation}\label{e4}
	\alpha_{2} = r_{2}^{3}  \left ( \frac{\epsilon_{2} - f_{1}\epsilon_{3} }{\epsilon_{2} + 2f_{1}\epsilon_{3} } \right), 
\end{equation} 
where $r_{2}$ is the nanoshell radius, $\epsilon_{2}$ is the permittivity of the shell, $\epsilon_{3}$ is the permittivity of the medium surrounding the shell, and 
\begin{equation}\label{e5}
	f_{1} = \left(1 - \frac{\alpha_{1}}{r_{2}^{3}} \right)\left(1 + 2 \frac{\alpha_{1}}{r_{2}^{3}} \right)^{-1}. 
\end{equation}

By inspection, there is a common pattern which $\alpha_{1}$ and $\alpha_{2}$ follow, and so do $f_{0}$ and $f_{1}$. We can utilize these patterns to obtain $\alpha_{3}$ and $f_{2}$ by induction, for a MNS with three concentric shells ($n = 3$), as follows: 
\begin{subequations} 
	\begin{align}
		\alpha_{3} & = r_{3}^{3}  \left ( \frac{\epsilon_{3} - f_{2}\epsilon_{4} }{\epsilon_{3} + 2f_{2}\epsilon_{4} } \right), \label{e6}\\
		f_{2} & = \left(1 - \frac{\alpha_{2}}{r_{3}^{3}} \right)\left(1 + 2 \frac{\alpha_{2}}{r_{3}^{3}} \right)^{-1}, \label{e7}
	\end{align}
\end{subequations}
where $\alpha_{3}$  is the dipole polarizability of the nanoshell, $\epsilon_{3}$ is the permittivity of the shell, $\epsilon_{4}$ is the permittivity of the medium surrounding the shell, and $r_{3}$ is the radius of the nanoshell. 
Hence, by induction, we can re-write Eqs. \eqref{e6} and \eqref{e7} for the $n$th shell to obtain:
\begin{subequations}
	\begin{align}
		\alpha_{n} & = 	r_{n}^{3}  \left ( \frac{\epsilon_{n} - f_{n-1}\epsilon_{n+1} }{\epsilon_{n} + 2f_{n-1}\epsilon_{n+1} } \right), \label{e8}\\
		f_{n-1} & = \begin{cases} 
			1, & n = 0\\
			\left(1 - \frac{\alpha_{n-1}}{r_{n}^{3}} \right)\left(1 + 2 \frac{\alpha_{n-1}}{r_{n}^{3}} \right)^{-1}, & n \ge 1
		\end{cases}\label{e9},
	\end{align} 
\end{subequations}
where $r_{n}$ is the radius of the $n$th nanoshell, $\epsilon_{n}$ 
is the permittivity of the $n$th shell, $\epsilon_{n+1}$ is the permittivity of the medium surrounding the $n$th shell, and 
$\alpha_{n-1}$ is the normalized (normalized by $4\pi\epsilon_{n}$) dipole polarizability of the $(n-1)$th shell, and 
$\alpha_{n}$ is the normalized (normalized by $4\pi\epsilon_{n+1}$) dipole polarizability of the $n$th shell.
An iterative algorithm proposed for calculating $\alpha_{n}$ can be found in Algorithm \ref{algo1} of the Appendix section.
Although substituting $n = 0, 1, 2,...$, in Eqs. \eqref{e8} and \eqref{e9} will reproduce $\alpha_{0}, \alpha_{1}, \alpha_{2}, ...$, 
we now need to use these equations to calculate the spectral properties of a given MNS, and compare our results to simulations, in order to ascertain their validity. One such property is absorption. In the LWA, the absorption efficiency of a MNS in the presence of an incident electric field can be calculated using the equation \cite{Luke22,Raza14,Herr16,BoHu08}: 
\begin{equation}\label{e10}
	Q_{abs} = \frac{k\Im[4\pi\epsilon_{n+1} \alpha_{n} ]}{\pi r_{n}^{2}}.
\end{equation}

\subsection{The Fr\"{o}hlich function}\label{ss2}
Given a MNP with a certain polarizability, the Fr\"{o}hlich condition \cite{Sam07,BoHu08,Radi20} states that the LSPRs supported by the MNP are the poles of the polarizability. In other words, the LSPRs correspond to frequencies where the real part of the denominator of the polarizability vanishes. 
Let us call this real part of the denominator of the polarizability the 
\emph{Fr\"{o}hlich function}. 
Starting with Eq. \eqref{a1}, the dipole polarizability of a spherical nanoshell with $n = 1$, the denominator of $\alpha_{1}$ simplifies to:
\begin{equation}\label{a12}
	D[ \alpha_{1} ] = 
	\epsilon_{1}\left[(\epsilon_{0} + 2\epsilon_{1}) + 2(r_{0}/r_{1})^{3}(\epsilon_{0} - \epsilon_{1})   \right]  
	+ 2 \left[(\epsilon_{0} + 2\epsilon_{1})- (r_{0}/r_{1})^{3}(\epsilon_{0} - \epsilon_{1})   \right] \epsilon_{2}.
\end{equation}
We can re-write Eq. \eqref{a12} as:
\begin{equation}\label{a13}
	D[ \alpha_{1} ] = 
	\epsilon_{1}D_{1}^{r}
	+ 2 N_{1}^{r} \epsilon_{2}.
\end{equation}
with 
\begin{subequations} 
	\begin{align}
		D_{1}^{r} & = D_{0} + 2N_{0}/r_{1}^{3}, \\
		N_{1}^{r} & = D_{0} - N_{0}/r_{1}^{3},
	\end{align}
\end{subequations}
and 
\begin{subequations} 
	\begin{align}
		D_{0} & = \epsilon_{0} + 2\epsilon_{1}, \\
		N_{0} & = r_{0}^{3}(\epsilon_{0} - \epsilon_{1}),
	\end{align}
\end{subequations}
so that the Fr\"{o}hlich function for the $n=1$ nanoshell is: 
\begin{equation}
	\mathcal{F}_{1} = \Re[ 	\epsilon_{1}D_{1}^{r}
	+ 2 N_{1}^{r} \epsilon_{2}  ].
\end{equation}
For the MNS with $n = 2$, the denominator of $\alpha_{2}$ given in Eq. \eqref{a6} simplifies to:
\begin{equation}\label{a14}
	D[ \alpha_{2} ] = 
	\epsilon_{2}[D_{1} + 2N_{1}/r_{1}^{3} ]  + 2[D_{1} - N_{1}/r_{1}^{3}  ] \epsilon_{3},
\end{equation}
with 
\begin{subequations} 
	\begin{align}
		N_{1} & = r_{1}^{3} \beta_{1}, \label{a15} \\
		D_{1} & = \varrho_{1}. \label{a16}
	\end{align}
\end{subequations}
We can re-write Eqs. \eqref{a15} and \eqref{a16} in terms of $N_{0}$ and 
$D_{0}$ to obtain:
\begin{subequations} 
	\begin{align}
		D_{1} & = \epsilon_{1}(D_{0} + 2N_{0}/r_{1}^{3}) + (D_{0} - N_{0}/r_{1}^{3})\epsilon_{2} = 
		\epsilon_{1}D_{1}^{r} + N_{1}^{r}\epsilon_{2} , \label{a17} \\
		N_{1} & = r_{1}^{3}[\epsilon_{1}(D_{0} + 2N_{0}/r_{1}^{3}) - (D_{0} - N_{0}/r_{1}^{3})\epsilon_{2}] = r_{1}^{3}[\epsilon_{1}D_{1}^{r} - N_{1}^{r}\epsilon_{2}], \label{a18}
	\end{align}
\end{subequations}
so that with $D_{2}^{r} = D_{1} + 2N_{1}/r_{1}^{3}$ and 
$N_{2}^{r} = D_{1} - N_{1}/r_{1}^{3}$, we obtain the Fr\"{o}hlich function for the $n=2$ nanoshell:
\begin{equation}\label{a19}
	\mathcal{F}_{2} = \Re[ 	\epsilon_{2}D_{2}^{r}
	+ 2 N_{2}^{r} \epsilon_{3}  ].
\end{equation}
Hence, by induction, the Fr\"{o}hlich function for the $n$th shell is: 
\begin{equation}\label{e11}
	\mathcal{F}_{n} = \Re[ \epsilon_{n}D^{r}_{n} + 2N^{r}_{n}\epsilon_{n+1} ].
\end{equation}
In Eq. \eqref{e11}, 
\begin{subequations}
	\begin{align}
		D^{r}_{n} & = \begin{cases} 
			1, & n = 0\\
			D_{n-1} + 2\frac{N_{n-1}}{r_{n}^{3}} , & n \ge 1
		\end{cases}\label{e8a},\\
		N^{r}_{n} & = \begin{cases} 
			1, & n = 0\\
			D_{n-1} - \frac{N_{n-1}}{r_{n}^{3}} , & n \ge 1
		\end{cases}\label{e8b},
	\end{align} 
\end{subequations}
where, for $n \ge 0$, 
\begin{subequations}
	\begin{align}
		D_{n} & =  \epsilon_{n}D^{r}_{n} + 2N^{r}_{n}\epsilon_{n+1} \label{e9a},\\
		N_{n} & =  r_{n}^{3}(\epsilon_{n}D^{r}_{n} -N^{r}_{n}\epsilon_{n+1}) \label{e9b}. 
	\end{align} 
\end{subequations}
The zeroes of the Fr\"{o}hlich function for the $n$th shell, 
i.e., frequencies (or excitation wavelengths) where $\mathcal{F}_{n} = 0$, 
correspond to the LSPRs supported by the MNS. 
An iterative algorithm proposed for calculating $\mathcal{F}_{n}$ can be found in Algorithm \ref{algo2} of the Appendix section.

\subsection{Simulations}\label{ss3}
The first set of simulations were based on the finite-element method (FEM). These FEM-based simulations were performed in the Wave Optics module of COMSOL Multiphysics$^{\circledR}$ software \cite{comsol} using a spherically-symmetric perfectly-matched layer (PML) and scattering boundary conditions applied in the internal PML surface. An $x-$polarized incident electric field, as described earlier, was applied to the MNS. The absorption efficiency of the nanoshell is calculated through the following expression \cite{Grand19,Luke22}:
\begin{equation}\label{e12}
	Q_{abs} = \frac{1}{P_{0}A}  \iiint P_{diss}dV . 
\end{equation}
The integral in Eq. \eqref{e12} is a volume integral of the total power dissipation density of the nanoshell, $P_{diss}$, where $P_{0}$ is the power density of the incident electric field, and $A$ is the area of the nanoshell obtained from a surface integral over the nanoshell surface \cite{Grand19}.

In the second set of simulations, we used scattnlay \cite{Pena17}---an open source code for investigating the scattering of electromagnetic radiation by a multilayered sphere---developed by Pal et al. \cite{Pena09,Pena13,Pena17} based on Mie theory (MT)\cite{BoHu08,Kerk69,Mie08}. 
In these MT-based simulations, the absorption efficiency of a given MNS is calculated from the extinction efficiency, $Q_{ext}$, and the scattering efficiency, $Q_{sca}$, using the following equation \cite{Pena09}:
\begin{equation}\label{e13}
	Q_{abs} = Q_{ext} - Q_{sca},
\end{equation}
where \cite{Pena09,Pena13}
\begin{subequations} 
	\begin{align}
		Q_{ext} & = \frac{2}{x^{2}_{n}}\sum_{l=1}^{\infty}(2l+1)\Re\Big(a_{l} +b_{l}\Big) , \label{e14} \\
		Q_{sca} & = \frac{2}{x^{2}_{n}}\sum_{l=1}^{\infty}(2l+1)\Big(|a_{l}|^{2} +|b_{l}|^{2}\Big) \label{e15}.
	\end{align}
\end{subequations}
In Eqs. \eqref{e14} and \eqref{e15}, $l = 1, 2,3, ...$ is an angular momentum number, denoting the number of multipoles over which the summation is done, $a_{l}$ and $b_{l}$ are scattering coefficients, and 
$x_{n}$ is the size parameter of the $n$th layer \cite{Pena09}.

As an example, we investigated nanoshells containing only one type of metal as well as one type of dielectric, as illustrated in Fig. \ref{f1}. To account for both intra- and inter-band electron damping in the metal, we used the Drude-Lorentz model for the permittivity of metals, given in Ref. \cite{Rakic98} as : 
\begin{equation}\label{e16}
	\varepsilon(\omega) = \varepsilon_{\infty} - 
	f_{0}\frac{\omega^{2}_{0}}{\omega(\omega-i\gamma_{0})}
	+\sum_{s=1}^{5}f_{s}\frac{\omega^{2}_{0}}
	{\omega^{2}_{s}-\omega^{2}+i\omega\gamma_{s}}, 
\end{equation}
where $\varepsilon(\omega)$ in the case of Fig. \ref{f1}, is the permittivity of the metallic core (or shell), $\omega$ is the frequency of the incident field, 
$\varepsilon_{\infty}$ is the high-frequency permittivity of the positive ion core in the metal, $f_{0}, \omega_{0}$, and $\gamma_{0}$ are the oscillator strength, plasma frequency, and damping rate of the intra-band electrons in the metal, respectively, $s$ is the number of Lorentz oscillators with oscillator strength $f_{s}$, damping rate $\gamma_{s}$, and frequency $\omega_{s}$, associated with inter-band electrons in the metal. The values of these parameters for the most common metals used in plasmonics are given in Ref. \cite{Rakic98}.
\begin{table}[h]
	\caption{List of the nanoshell configurations we investigated. The letter 
	\textquotedblleft D" stands for dielectric and \textquotedblleft M" stands for metal. The starting configurations differ but we have kept the motif uniform. For the nanoshells in Fig. \ref{f1}($a$), the number of layers, $n$, are odd, i.e., $n = 1, 3, 5$, while for the nanoshells in Fig. \ref{f1}($b$), the number of layers, $n$, are even, i.e., $n = 2, 4, 6$.}\label{t1}
		\begin{tabular}{@{}lllll@{}}
			\toprule
			Number of layers & MNS& Innermost core & Starter & Motif\\ \midrule
			1                & DM & D & DM &  \\ 
			2                & MDM & M & MDM  &  \\ 
			3              & DMDM  & D  & DM &  \\ 
			4               & MDMDM & M & MDM & DM \\ 
			5              & DMDMDM  & D & DM &  \\ 
			6               & MDMDMDM & M &MDM &  \\ 
			\botrule
		\end{tabular}
\end{table}

\section{Results and Discussion}\label{sec3}
Table \ref{t1} shows the material compositions of the MNS we investigated using the proposed dipole polarizability formula. Here, we will consider glass, with a permittivity of $2.25$ as the dielectric as well as the medium surrounding the outermost nanoshell, and gold, modelled with the Drude-Lorentz parameters given in Ref. \cite{Rakic98}, as the metal. We chose to use only two materials for simplicity 
and ease of analysis of the results
but the formula is also valid when different material compositions are used. The radius of the core of the nanoshell in each of the nanoshell configurations in Table \ref{t1} (or in Fig. \ref{f1}) is kept constant at $15$ nm. The shells were modelled as equidistant shells, each of thickness, $5$ nm. 

\subsection{Absorption spectra}
\begin{figure}[t]
	\centering 
	\includegraphics[width = .46\textwidth]{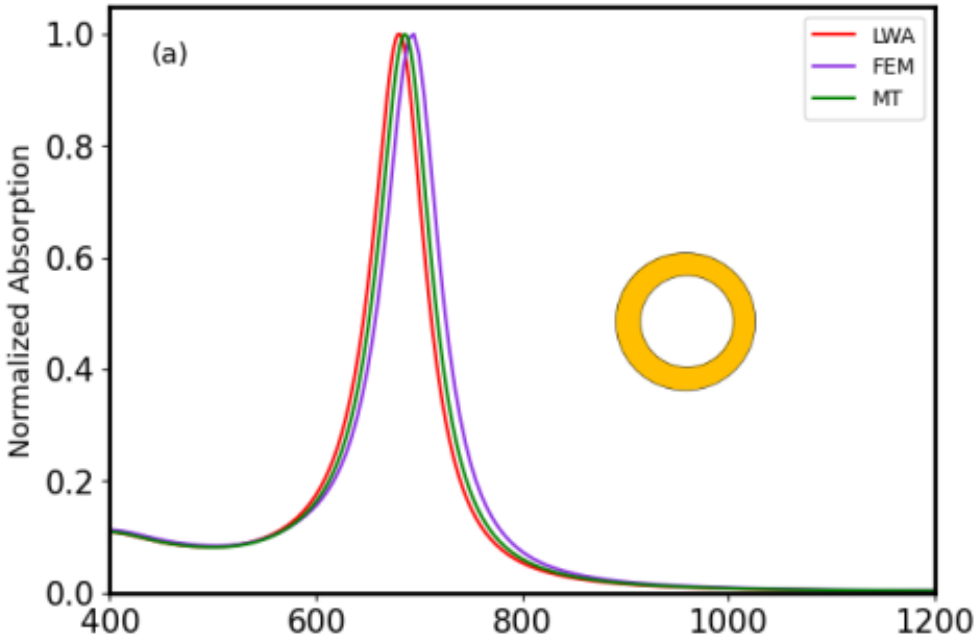}
	\includegraphics[width = .44\textwidth]{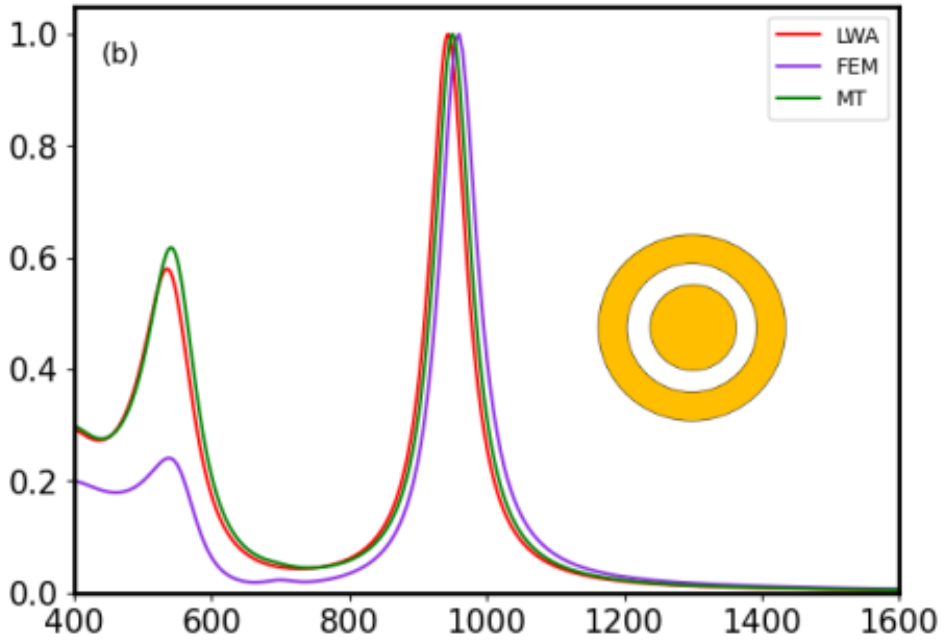}\vspace{0.2cm}\\
	\includegraphics[width = .46\textwidth]{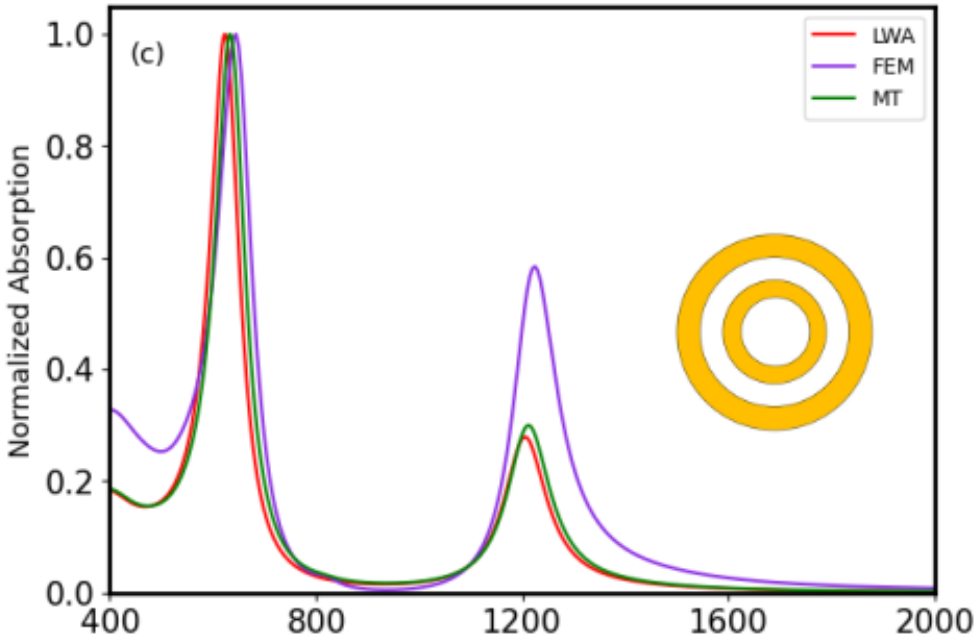}
	\includegraphics[width = .44\textwidth]{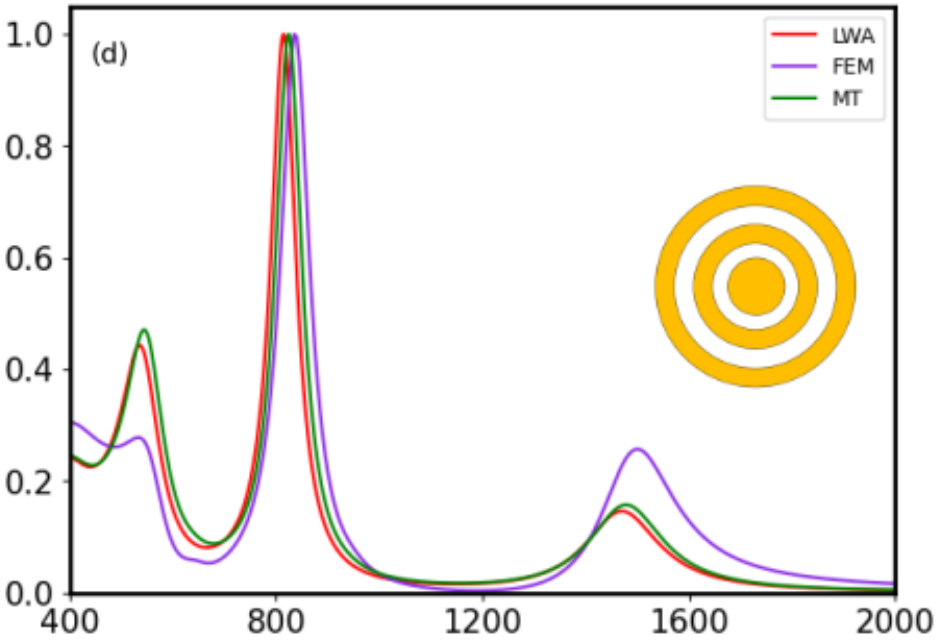}\vspace{0.2cm}\\
	\includegraphics[width = .46\textwidth]{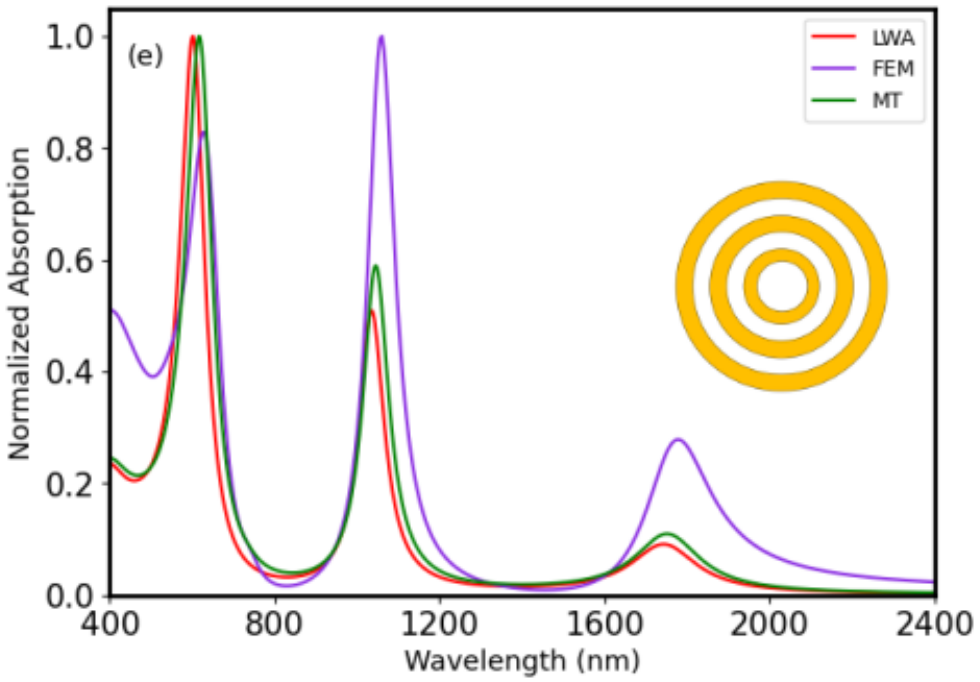}
	\includegraphics[width = .44\textwidth]{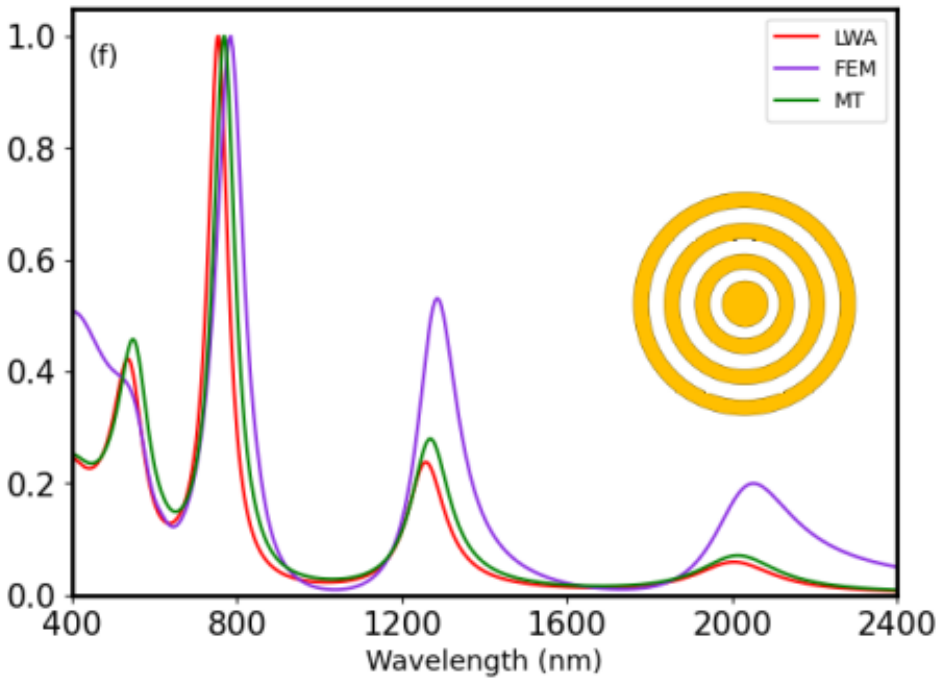}
	\caption{Normalized absorption efficiencies (normalized by the maximum value of $Q_{abs}$ in each data set) of the MNS. Nanoshells with: 
		(a) $n = 1$, (b) $n = 2$, (c) $n = 3$, (d) $n = 4$, (e) $n = 5$, and (f) $n = 6$. Our approach (via LWA): Red curves. COMSOL Multiphysics (via FEM): Blueviolet curves. scattnlay (via MT): Green curves. 
	}\label{f2}
\end{figure}
Using the MT-based results as the benchmark---since it is based on exact multipole expansion of the field intensities in the core, shell, and host mediums of the nanoshell--- 
the usual limitations of the LWA, i.e., under-prediction of the absorption efficiencies (and other efficiency factors \cite{Raza14}) and under-prediction of the LSPR shift \cite{Raza14,Wub14} (less redshift), can be seen in Fig. \ref{f2}. However, in the MNS configurations in Fig. \ref{f2}, the LWA does not under-predict all the absorption peaks---it appears that one of the absorption peaks is always in good agreement with the MT-based simulations. 
For instance, as can be seen in Fig. \ref{f2}, their absorption peaks are comparable at: 693 nm (for $n=1$), 644 nm (for $n=3$), and 628 nm (for $n=5$), for the MNS with a dielectric core, and at: 958 nm (for $n=2$), 835 nm (for $n=4$), and 784 nm (for $n=6$), for the the MNS with a metallic core---under-going a blueshift as the number of shells increases. 
The FEM-based simulations also follow a similar trend as the LWA except that it over-predicts some of the absorption peaks. 
Though the most redshifted peaks 
(Figs. \ref{f2}(e) and \ref{f2}(f)) are more suppressed in the MT-based results, the number of peaks are always the same when compared to the LWA results, unlike the FEM-based simulations, 
showing that the LWA model is in good agreement with the MT-based simulations. 

On the other hand, when compared to the MT-based simulations, the FEM-based simulations are very accurate in predicting the LSPR shift, as shown by the agreement in the wavelengths corresponding to the absorption peaks in Fig. \ref{f2}. 
This is because the FEM-based simulations account for phase retardation effects between oscillating modes of the scattered fields in the MNP \cite{Grand19}. 
We therefore attribute the under-prediction of the LSPRs by the LWA to mostly retardation effects.  
The difference in the absorption peaks of the FEM-based and MT-based simulations most likely stems from the approximation of the spherical Bessel functions in the former while the latter uses the exact functions. It is likely that these approximations accumulate as the number of shells increase causing a noticeable discrepancy between the two simulations. 
The accuracy of the FEM-based simulations depends largely on the properties of the mesh elements used as well as on the PML features, so that in principle, the simulation can be improved. However, this work is the first to compare comsol multiphysics and scattnlay simulations, using MNSs as the target nanostructure, to the best of our knowledge. 

As we will show in Section \ref{s3ss2} using near-field intensity plots, the LSPRs in Fig. \ref{f2} (green curves) all have a dipole character, which means that their origin is most likely to have been dominated by dipolar modes, due to the small particle size of the nanoshells we studied. 
In addition, Eq. \eqref{e8}, which we used to calculate the absorption spectra of the MNS in Fig. \ref{f2} (red curves) is a dipole polarizability. Thus, the LSPRs in the LWA spectra are all due to dipole hybridization between the core and shell plasmons of the MNS. Hence, in the LWA, these LSPRs all have a dipole character. 

A noticeable difference between the absorption spectra of the nanoshells with a dielectric core (Figs. \ref{f2}(a), (c), and (e) with $n = 1,3$ and $5$, respectively) and those with a metallic core (Figs. \ref{f2}(b), (d), and (f), with $n = 2, 4$ and $6$, respectively), is the enhancement versus suppression of the leading absorption peak. This can be attributed to the presence of cavity plasmons in the core (in the case of Figs. \ref{f2}(a), (c), and (e)) and solid plasmons in the core (in the case of Figs. \ref{f2}(b), (d), and (f)). In the MNS in Figs. \ref{f2}(a), (c), and (e), hybridization of the cavity plasmons of the core with the solid plasmon of the innermost shell results in the formation of the leading LSPR, while in the MNS in Figs. \ref{f2}(b), (d), and (f)), the solid plasmons of the core hybridize with the cavity plasmon of the innermost shell to form the leading LSPR. 
Previous studies have reported similar trends for $n = 2$ \cite{Luke22,Pena13} and $n = 3$ \cite{Pena13,Far14}. 
As we will discuss in the next section, any one of the LSPRs supported by the MNS is either due to a bonding mode, an anti-bonding mode, or a non-bonding mode \cite{Prod03,Prod04,Park09,Ma17}.

\subsection{Plasmon resonances}\label{s3ss2}
The LSPRs supported by the $n$th nanoshell can be further investigated using graphical solutions of the Fr\"{o}hlich function given in Eq. \eqref{e11}. Electric field enhancement plots produced from simulations in scattnlay \cite{Pena17} reveal the hybridized nature of these LSPRs, in agreement with plasmon hybridization theory (PHT) \cite{Prod03}. We will also characterize the LSPRs as either due to a bonding, an anti-bonding, or a non-bonding dipole mode, using the behaviour of the electric field distributions in the core and shell regions of the MNS. 

\begin{figure}[t]
	\centering 
	\includegraphics[width = 1\textwidth]{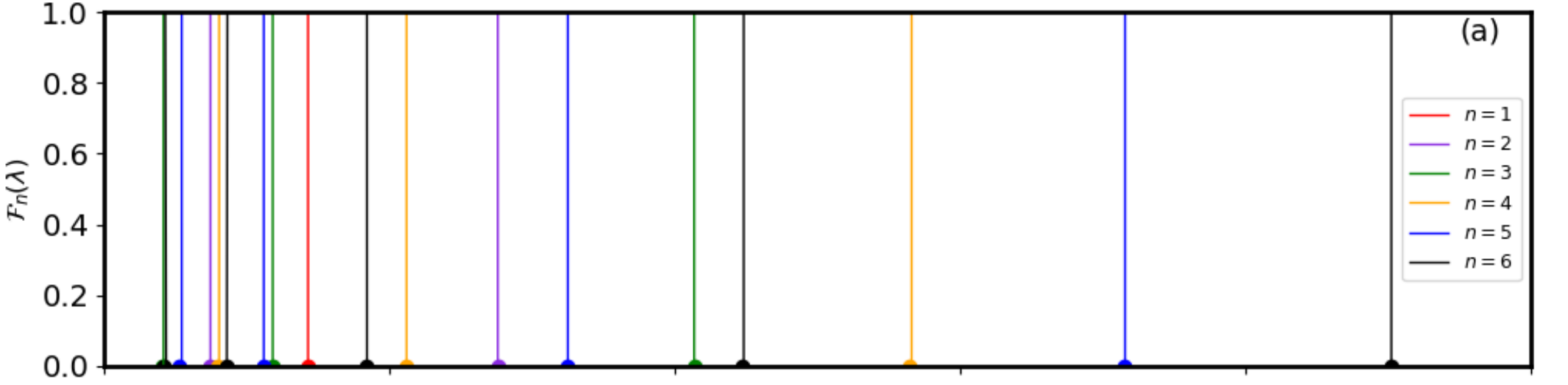}\\
	\includegraphics[width = 1\textwidth]{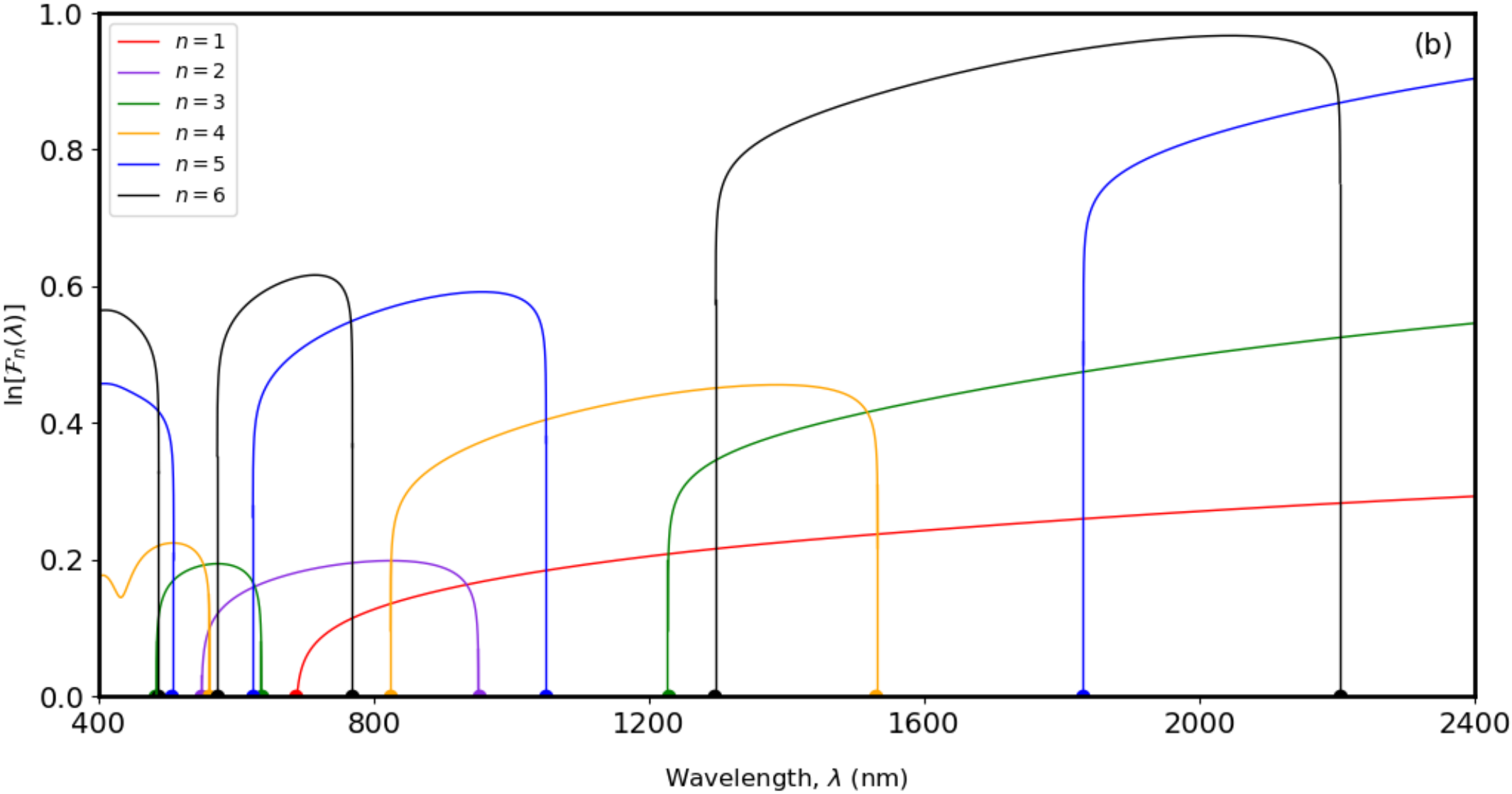}
	\caption{
		(a) A plot of the Fr\"{o}hlich function showing the plasmon spectrum of the MNSs. The LSPRs correspond to wavelengths at which $\mathcal{F}_{n}(\lambda) = 0$. 
		The Fr\"{o}hlich function is either an $n$ degree polynomial or an $(n-1)$ degree polynomial in $\lambda$. The plot has been zoomed-in to show only the parts of the polynomials that pass through $\mathcal{F}_{n}(\lambda) = 0$. (b) Natural logarithm of the Fr\"{o}hlich function, $\ln[\mathcal{F}_{n}(\lambda)]$, showing a more detailed version of the plasmon spectrum of the MNSs. The paired and/or unpaired LSPRs are respectively due to the convex and/or concave curves in the Fr\"{o}hlich function of each MNS.
	}\label{f3}
\end{figure}

Fig. \ref{f3}(a) shows the plasmon spectrum of the MNS produced by plotting the Fr\"{o}hlich function. 
There are more LSPRs located at shorter wavelengths (between 400 nm and 1200 nm) than at longer wavelengths (between 1200 nm and 2400 nm), which can be attributed to particle size effects. The number of LSPRs increase with increase in the number of metallic shells in the MNS. This is due to an increase in the number of wavelength-dependent permittivity terms in the polarizability. 
Also, the spacing between the LSPRs increases as the number of shells increases. 
For $n=1, 2$, and $n=4$, the number of LSPRs predicted by the Fr\"{o}hlich function agrees with the absorption spectra (Figs. \ref{f2}(a), (b) and (d), respectively). However, for $n = 3, 5$, and $6$, respectively, the Fr\"{o}hlich function predicts an extra LSPR in each case compared to the absorption spectra (Figs. \ref{f2}(c), (e), and (f), respectively). This is most likely due to the dipole moment of the extra LSPR being too weak to contribute to the absorption spectra in Fig. \ref{f2}, making it to appear invisible in the spectra. According to Figs. \ref{f3}(a) and (b), the extra LSPR is located at a shorter wavelength than the leading LSPR in the absorption spectra in Figs. \ref{f2}(c), (e), and (f), respectively.  

\begin{figure}[h]
	\centering 
	\includegraphics[width = 0.97\textwidth]{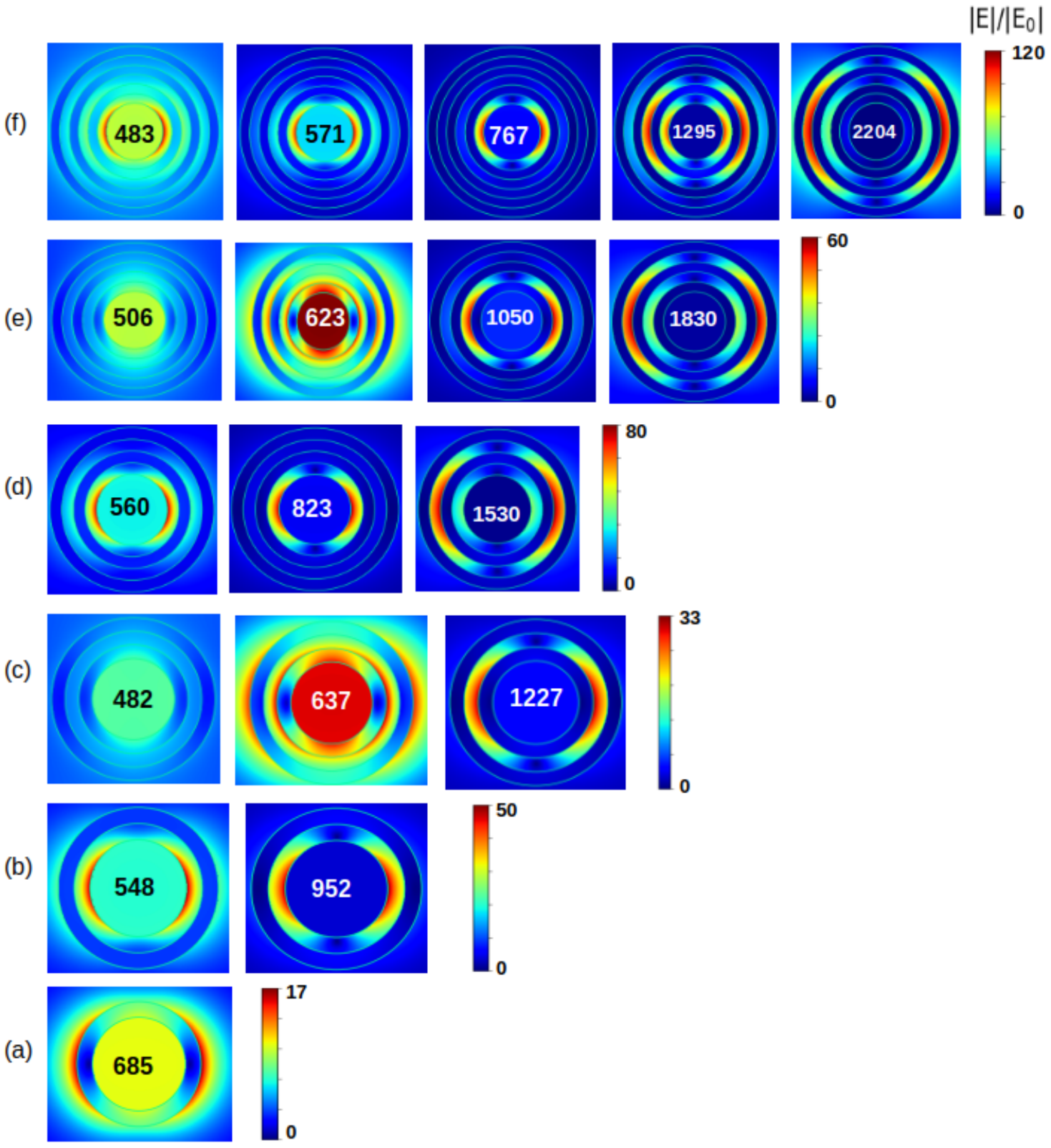}
	\caption{Contour plots of the electric field enhancement, $|\mathrm{E}|/|\mathrm{E}_{0}|$, inside and outside the MNS, in the xy-plane, using scattnlay \cite{Pena09,Pena17}, where $|\mathrm{E}|$ is the magnitude of the incident and scattered fields and $|\mathrm{E}_{0}|$ is the magnitude of the incident field.
		The contour plots were obtained at the LSPRs supported by each MNS in the LWA via the Fr\"{o}hlich function: (a) $n = 1$, (b) $n=2$, (c) $n = 3$, (d) $n=4$, (e)
		$n=5$, and (f) $n=6$. These LSPRs are indicated (in nm units) in the core of each nanoshell.
	}\label{f4}
\end{figure}

The $n = 1$ nanoshell supports two LSPRs---a bonding dipole (BD) LSPR and an anti-bonding dipole (ABD) LSPR, according to PHT \cite{Prod03,Prod04}.
The BD LSPR ($\omega_{-}$) is a longer-wavelength mode formed due to the symmetric coupling between solid sphere plasmons of the shell and cavity sphere plasmons of the core while
the ABD LSPR ($\omega_{+}$) is a shorter-wavelength mode formed as a result of antisymmetric coupling between cavity sphere plasmons of the core and the solid sphere plasmons of the shell. 
In the BD mode, solid plasmons contribute more than cavity plasmons while the opposite is the case in the ABD mode \cite{Prod04}. 
However, the dipole moment of the ABD mode is very weak, and therefore, it can only be visible in spectral calculations when inter-band damping is ignored \cite{Prod04,Park09}, 
which is not the case in this work. 
Fig. \ref{f4}(a) shows that the LSPR at $685$ nm is the BD mode since the incident electric field experiences the most enhancement outside the nanoshell in contrary to the ABD mode, where the electric field has been shown to be more enhanced inside the nanoshell (see Ref. \cite{Park09}, Fig. 1). 
Hence, the LSPR in Fig. \ref{f3} (red line/curve), for $n = 1$, is a BD mode.

The $n=2$ nanoshell supports three LSPRs --- a BD LSPR, an ABD LSPR, and a non-bonding dipole (NBD) LSPR \cite{Bard10,Ma17}. 
The NBD LSPR ($\omega_{+}^{-}$) is a short-wavelength mode formed by symmetric coupling between the solid sphere plasmon of the core and the ABD mode ($\omega_{+}$) of the nanoshell. Due to its weak dipole moment, it is only visible in the spectra of silver nanoshells \cite{Nia14}, where plasmon damping is significantly lower, or in the absence of inter-band effects. Hence, in our case, the two LSPRs supported by the $n=2$ MNS, as shown in Fig. \ref{f3}, are an ABD mode ($\omega_{-}^{+}$) and a BD mode ($\omega_{-}^{-}$). 
The longer-wavelength, BD ($\omega_{-}^{-}$) mode, is due to anti-symmetric coupling between the BD ($\omega_{-}$) nanoshell plasmon and the solid sphere plasmon of the core, while 
the shorter-wavelength, ABD ($\omega_{-}^{+}$) mode, is due to symmetric coupling between the BD ($\omega_{-}$) nanoshell plasmon and the solid sphere plasmon of the core \cite{Bard10,Nia14,Ma17}.
As shown in Fig. \ref{f4}(b), for $n = 2$, the enhanced electric field is situated entirely in the dielectric shell at $952$ nm (the BD mode) compared to the electric field distribution at $548$ nm (the ABD mode), in agreement with previous work \cite{Ma17}.

In Fig. \ref{f3}(b), a plot of the natural logarithm of the 
Fr\"{o}hlich function is shown.
Here, the positions of the LSPRs are much more visible, and the plot also reveals an intricate detail about the plasmonic behaviour of the LSPRs---the pairing of certain LSPRs beyond $n=1$ due to the convex curves in the Fr\"{o}hlich function. 
We propose that this pairing behaviour can be used in determining whether a given LSPR supported by the MNS has a BD or an ABD character.
For instance, when $n=2$, we have the blueviolet curve in Fig. \ref{f3}(b) which
contains a pair of LSPRs corresponding to an ABD mode at $548$ nm and a BD mode at $952$ nm, in agreement with  Fig. \ref{f2}(b).
However, we will use the rest of this section to show that 
any given pair of LSPRs in the Fr\"{o}hlich function does not always consist of an ABD and a BD mode. On the other hand, 
subsequent analysis of the unpaired LSPRs, shows that 
a shorter-wavelength LSPR (the leftmost LSPR in the Fr\"{o}hlich function) that does not participate in the pairing behaviour is always an ABD mode, while a longer-wavelength LSPR (the rightmost LSPR in the Fr\"{o}hlich function) that does not form a pair has a BD character. These results are summarized in Table \ref{t2}.

To support the above claim, consider the rest of the electric field intensity plots shown in Figs. \ref{f4}((c)--(f)). 
The $n=3$ MNS supports up to four LSPRs in the absence of inter-band damping in gold, according to PHT \cite{Prod04}. 
These LSPRs are formed as a result of hybridization between the inner- and outer-nanoshell BD and ABD modes. They include: 
$\omega_{-}^{-}$, a BD mode due to symmetric coupling between the inner- and outer-nanoshell BD modes, 
$\omega_{-}^{+}$, an ABD mode due to anti-symmetric coupling between the inner- and outer-nanoshell BD modes,
$\omega_{+}^{-}$, a NBD mode due to anti-symmetric coupling between the inner- and outer-nanoshell ABD modes, and $\omega_{+}^{+}$, a NBD mode due to anti-symmetric coupling between the inner- and outer-nanoshell ABD modes. The dipole moment of the $\omega_{+}^{+}$ mode is very weak, and thus, it can only be visualized with a Drude model for gold. Likewise, the $\omega_{+}^{-}$ mode is also a dark mode, since its contribution is not visible in the absorption spectra in Fig. \ref{f2}(c), but it does appear in the Fr\"{o}hlich function at $482$ nm (green line/curve in Fig. \ref{f3}). The weak-dipole moment explanation for the absence of 
$\omega_{+}^{-}$ (NBD) mode in Fig. \ref{f2}(c) is also supported by Fig. \ref{f4}(c), where the electric field is weakly distributed in the core and inside the second shell at $482$ nm. 
On the other hand, the $\omega_{-}^{+}$ and $\omega_{-}^{-}$  modes are visible in Fig. \ref{f2}(c) due to their large dipole moments. This gave rise to the electric field distributions shown in Fig. \ref{f4}(c), where at $637$ nm, 
the electric field is distributed in both the dielectric core and metallic shells, compared to the field distribution at
$1227$ nm, where the field is only enhanced inside one of the dielectric shells. Therefore, compared to Fig. \ref{f3} (green curve), we have a LSPR pair corresponding to an ABD mode at $482$ nm and another ABD mode at $637$ nm, as well as a rightmost LSPR corresponding to the BD mode at $1227$ nm. 

To characterize the rest of the LSPRs in Figs. \ref{f2} and \ref{f3}, we utilized the electric field enhancement plots in 
Fig. \ref{f4}((d) -- (f)). 
Using the same analogy as above, for $n = 4$ (Fig. \ref{f4}((d)),
we assigned the LSPRs at both $823$ nm and $1530$ nm a BD character, since the enhanced electric field is mostly distributed in the dielectric shells, i.e., the electric field enhancements in the metallic regions are negligible compared to the LSPR at $560$ nm, which we assigned an ABD character.
Thus, for $n=5$, the LSPRs at both $1050$ nm and $1830$ nm were assigned a BD character while the LSPRs at both $506$ nm and $623$ nm, where some significant field enhancement exist in the metallic regions, were assigned an ABD character. 
Similarly for $n=6$, the LSPRs at $767$ nm, $1295$ nm, and $2204$ nm were assigned a BD character, while those at both $483$ nm and $571$ nm were assigned an ABD character. 
As shown in Table \ref{t2}, a LSPR pair located in the short-wavelength region of the plasmon spectrum is most likely an ABD-ABD pair, while a pair located in the long-wavelength region consists of a BD-BD pair. However, a LSPR pair in-between the short- and long-wavelength regions is most likely an ABD-BD pair.
\begin{table}[h]
		\caption{List of the nanoshell configurations we investigated and the LSPRs they support, according to the Fr\"{o}hlich function. Here, \textquotedblleft~ -- \textquotedblright ~stands for a LSPR pair in the plasmon spectrum.}\label{t2}
		\begin{tabular}{@{}lll@{}}
			\toprule
			Number of shells & MNS& LSPRs \\ 
			\midrule
			1                & DM & 685 nm (BD)  \\
			2                & MDM & 548 nm (ABD) -- 952 nm (BD)   \\ 
			3              & DMDM  & 482 nm (ABD) -- 637 nm (ABD), 1227 nm (BD)  \\ 
			4               & MDMDM & 560 nm (ABD), 823 nm (BD) --  1530 nm (BD)  \\ 
			5              & DMDMDM  & 506 nm (ABD), 623 nm (ABD) -- 1050 nm (BD), 1830 nm (BD)   \\ 
			6               & MDMDMDM & 483 nm (ABD), 571 nm (ABD) -- 767 nm (BD),\\
			              &  & 1295 nm (BD) -- 2204 nm (BD)\\
			\botrule
		\end{tabular}
\end{table}

\section{Conclusion}
The dipole polarizability formula we have proposed for spherical MNSs is straightforward to implement. It is an alternative for obtaining the dipole polarizability of MNSs in the LWA without going through the conventional approach based on scattering coefficients. 
The formula works well especially when used to predict the LSPRs supported by a MNS via the proposed Fr\"{o}hlich function. We have shown that the LSPRs correspond to the zeroes of the Fr\"{o}hlich function for the $n$th shell, 
and that a pairing behaviour by some of the LSPRs identified in the Fr\"{o}hlich function might be useful for mode characterization. However, we maintain that it is an approximate formula strictly valid in the LWA, and should not be used otherwise. The limit of validity of the formula was revealed by comparison with numerical simulations, showing that it reproduces the usual LWA results. 
Therefore, for the analytical modelling of the optical response of MNSs, the proposed formula is neither meant to replace the standard non-LWA approaches nor to eliminate the limitations of the LWA, but rather to highlight its simplicity and applicability when compared to conventional LWA. 

\section*{Funding}
This research was supported financially by the Department of Science and Innovation (DSI) through the South African Quantum Technology Initiative (SA QuTI), Stellenbosch University (SU), the National Research Foundation (NRF), and the Council for Scientific and Industrial Research (CSIR). 

\section*{Code Availability} 
Scattnlay codes, and Python codes written for this work are available on request via the corresponding author's email. 

\section*{Appendix: Algorithms for calculating the electrostatic dipole polarizability and Fr\"{o}hlich function of the nth shell}\label{secA1}
\begin{algorithm}
	\caption{Electrostatic dipole polarizability of the $n$th shell}\label{algo1}
	\begin{enumerate}
	\item Create a list of $r_{n}$, e.g., for $n = 6, r_{n} = [r_{0}, r_{1}, r_{2}, r_{3}, r_{4}, r_{5}, r_{6}   ]$ 
\item Create a list of $\epsilon_{n}$, e.g., for $n = 6,  \epsilon_{n} = [\epsilon_{0}, \epsilon_{1}, \epsilon_{2}, \epsilon_{3}, \epsilon_{4}, \epsilon_{5}, \epsilon_{6} ]$ 
\item Create a list of $\epsilon_{n+1}$, e.g., for $n = 6,  \epsilon_{n+1} = [\epsilon_{1}, \epsilon_{2}, \epsilon_{3}, \epsilon_{4}, \epsilon_{5}, \epsilon_{6}, \epsilon_{7} ]$ 
\item Calculate the dipole polarizability, $\alpha_{0}$, of the core 
\item Calculate $f_{n-1}$ and $\alpha_{n}$ for the first shell, $n = 1$
\item Update the dipole polarizability, i.e., set $\alpha_{0} = \alpha_{1}$  
\item Repeat steps 5 and 6 for subsequent shells until the $n$th shell
\item Return $\alpha_{0}$	
	\end{enumerate}
\end{algorithm}
\begin{algorithm}
	\caption{The Fr\"{o}hlich function of the $n$th shell}\label{algo2}
\begin{enumerate}
	\item Create a list of $r_{n}$, e.g., for $n = 6, r_{n} = [r_{0}, r_{1}, r_{2}, r_{3}, r_{4}, r_{5}, r_{6}   ]$ 
	\item Create a list of $\epsilon_{n}$, e.g., for $n = 6,  \epsilon_{n} = [\epsilon_{0}, \epsilon_{1}, \epsilon_{2}, \epsilon_{3}, \epsilon_{4}, \epsilon_{5}, \epsilon_{6} ]$
	\item Create a list of $\epsilon_{n+1}$, e.g., for $n = 6,  \epsilon_{n+1} = [\epsilon_{1}, \epsilon_{2}, \epsilon_{3}, \epsilon_{4}, \epsilon_{5}, \epsilon_{6}, \epsilon_{7} ]$ 
	\item Calculate $N_{0}$ and $D_{0}$
	\item Calculate $D_{1}^{r}$ and $N_{1}^{r}$
	\item Calculate $\mathcal{F}_{1}$
	\item Calculate $D_{1}$ and $N_{1}$
	\item Update $D_{0}$ and $N_{0}$, i.e., set $N_{0} = N_{1}$ and $D_{0} = D_{1}$
	\item Return $\Re[ \mathcal{F}_{1}  ]$
	\item Repeat steps 5 to 9 for subsequent shells until the $n$th shell
\end{enumerate}
\end{algorithm}

\bibliography{Bibliography} 
\end{document}